\documentclass[twocolumn, longbibliography, nofootinbib]{revtex4-2}

\usepackage{amsmath}
\usepackage{amssymb}
\usepackage[dvipsnames]{xcolor}
\usepackage{hyperref}
\hypersetup{
	colorlinks=true,
	linkcolor=blue,   
	urlcolor=blue,
}
\usepackage{graphicx}

\newcommand{\pardag}{\partial^\dagger\!}
\renewcommand{\cal}{\mathcal}
\newcommand{\sys}{\text{sys}}
\newcommand{\bath}{\text{bath}}
\newcommand{\Z}{\mathbb{Z}}

\begin{document}

\title{Self-correction from higher-form symmetry protection on a boundary}
\author{Charles Stahl}
\affiliation{Department of Physics and Center for Theory of Quantum Matter, University of Colorado, Boulder, CO 80309, USA}

\begin{abstract}

Recent work has shown that a self-correcting memory can exist in 3 spatial dimensions, provided it is protected by a 1-form symmetry. Requiring that a system's dynamics obey this type of symmetry is equivalent to enforcing a macroscopic number of symmetry terms throughout the bulk. In this paper, we show how to replace the explicit 1-form symmetry in the bulk with an emergent 1-form symmetry. Although the symmetry still has to be explicitly enforced on the boundary, this only requires $\cal{O}(L^2)$ terms instead of $\cal{O}(L^3)$ terms. We then reinterpret this boundary as a symmetry-protected topological defect in a bulk topological order. Defects can have interesting memory properties even in the absence of symmetry.

\end{abstract}
	
\date{\today}
	
\maketitle


\section{Introduction} \label{sec:intro}

Systems that can store quantum information for an extended period of time while interacting with a noisy environment will be integral components in any scalable implementation of quantum computation. Two important classes of such systems are \emph{fault-tolerant} quantum memories and \emph{self-correcting} quantum memories. From the perspective of condensed matter physics, fault-tolerant quantum memories can store quantum information indefinitely while evolving at zero temperature in the thermodynamic limit, even in the presence of small perturbations. The paradigmatic example is the two-space-dimensional (2d) toric code~\cite{Kitaev2003}, which is \emph{topologically ordered}. Topologically ordered systems~\cite{Wen1990} can generically provide fault tolerance by storing quantum information in their space of degenerate ground states. On the other hand, self-correcting quantum memories must be able to store quantum information indefinitely even at some nonzero temperatures, again in the thermodynamic limit. While the 4d toric code~\cite{Dennis2002} is self-correcting, there are no known examples in three dimensions. The 4d toric code remains topologically ordered in the temperature range in which it is self-correcting~\cite{Hastings2011}, but there is no general mathematical  result on the connection between self-correction and finite-temperature topological order~\cite{RobertsBartlett}. The existence of robust self-correction in 3d remains an open question.

The hunt for self-correction in systems with generic dynamics has generated interesting physics even where it hasn't achieved its central goal. In 2d, the introduction of exactly-solvable models has shed light on the classification of topological phases. In 3d, a direct search over a space of models~\cite{HaahCode} did not result in self-correction~\cite{Siva2017, PremHaahNandkishore}, but did kick off the study of fractons~\cite{Vijay2016, Nandkishore2019, Pretko2020}. 

In a parallel line of development, progress has been made by coupling toric codes to 2d~\cite{Hamma2009} or 3d~\cite{Pedrocchi2013} bosons. In the presence of diverging couplings (for the 2d bosons) or fine-tuned dynamics (for both), the bosonic systems can restore self-correction to the toric code. Generic perturbations destabilize the memory properties~\cite{LandonCardinal2015}. 

A more recent model from Roberts and Bartlett~\cite{RobertsBartlett} achieves a similar result by coupling to a bulk lattice spin model, with the advantage that bulk local Hilbert spaces are finite-dimensional. The model still requires fine-tuned dynamics, but encodes the fine-tuning in a \emph{higher-form symmetry}. Higher-form symmetries~\cite{Nussinov2009, Gaiotto2015, LakeHigher} are local symmetries that are not gauge symmetries, in that states related by a symmetry transformation are not identified as physically equivalent. The locality of the symmetry means that requiring the dynamics to respect the symmetry is a very strong constraint. For the model in Ref.~\cite{RobertsBartlett}, the dynamics must respect a number of constraints that scales with the volume of the system. We will refer to the model as the Roberts-Bartlett model.

In this paper, we show that the bulk higher-form symmetry need not be enforced. Instead, we only need to enforce the symmetry on the boundary.
First, in Sec.~\ref{sec:back} we review the physics of the existing models of quantum memories. Along the way, we look for insight into what role the higher-form symmetry plays in self-correction in the Roberts-Bartlett model. Then, in Sec.~\ref{sec:boundary} we construct a new model that only requires a higher-form symmetry to be enforced on the boundary. The new model has a topologically ordered bulk, so there is an emergent bulk higher-form symmetry~\cite{WenHigher}, which need not be enforced. We discuss the physical interpretation of the new model as a topological defect in Sec.~\ref{sec:interp}.
Finally, we ponder some possible future directions in Sec.~\ref{sec:conc}.

\section{Background} \label{sec:back}

Here we will review the physics that will be useful in motivating and understanding the model presented in the next section. First we will focus on various quantum memories and behavior at temperatures above zero. That material is reviewed thoroughly in Ref.~\cite{Brown2016}. We next define higher-form symmetries, leaning on the toric code for interpretation. Then, we motivate the power of higher-form symmetries for quantum memories and introduce the Roberts-Bartlett model.

\subsection{Quantum memories and nonzero temperature} \label{sub:mems}

Quantum memories store quantum information for an extended period of time by using special protected states. We call the states logical states, and the operators that act within the space of logical states are logical operators. The error model that acts on the quantum memory can not apply logical operators, possibly up to some probability cutoff or timescale. 
In this sense quantum memories are a generalization of quantum error-correcting codes, which are in turn a generalization of classical error correcting codes.

In this paper we will focus on Hamiltonian lattice models, where the logical states are ground states of the system. The first such system that was studied as a quantum memory was the toric code~\cite{Kitaev2003}. This is a 2d lattice model, with qubits living on the edges of the lattice. The Hamiltonian is 
\begin{align}
\begin{aligned}
H_\text{TC} &= -\sum_v A_v - \sum_f B_f,\\
A_v &= \prod_{e\in\partial^\dag v} X_e,\qquad B_f = \prod_{e \in \partial f} Z_e, \label{eq:HTC}
\end{aligned}
\end{align}
where $\partial f$ is the four edges in the boundary of the face $f$, and $\partial^\dag v$ is the four edges that form a ``star": the dual boundary of the vertex $v$.  We use $X$ and $Z$ to denote the Pauli matrices $\sigma^x$ and $\sigma^z$, respectively.

All $A_v$ and $B_f$ terms commute with each other. The only time is this is not obvious is when the vertex $v$ is on the boundary of the face $f$. In that case, $A_v$ and $B_f$ share two edges and the signs from commuting two $X$ operators past two $Z$ operators cancel, so the full terms commute. Since all terms in the Hamiltonian can be simultaneously satisfied, the model is exactly solvable. Ground states are $+1$ eigenstates of all $A_v$ and $B_f$ operators. Carefully counting the degrees of freedom and the ground state constraints shows that, while the constraints locally use up all the degrees of freedom, there are some global degrees of freedom that are not constrained, giving degenerate ground states.
The ground state degeneracy depends on the topology of the manifold on which the lattice is placed. 

The spectrum of the toric code contains anyons, or topologically charged excitations. The topological charge means that the anyons have nontrivial Aharonov-Bohm phases with respect to each other. Incomplete logical operators create and remove anyons at their endpoints, or  transport anyons. Complete logical operators (defined on topologically nontrivial closed strings) tunnel anyons across the system in non-contractible loops.  Even when perturbations are introduced to~\eqref{eq:HTC}, the tunneling amplitude is exponentially small in system size. As a result, the system can evolve under its own dynamics for a time $\tau$ without losing the stored information. As long as the system is at absolute zero temperature, the memory time $\tau$ diverges in the thermodynamic limit, up to some critical perturbation strength~\cite{Dennis2002}. As we said before, this is the defining feature of fault-tolerant quantum memories.

The toric code and related systems possess topological order~\cite{Wen1990}, a type of order with no local order parameter. Instead, the different ground states can only be distinguished by order parameters that are topologically nontrivial. In fact, the order parameters are the previously-mentioned logical operators. Due to the absence of local order parameters, the topological order is robust to small perturbations. By this we mean that the distinct ground states remain degenerate, up to corrections that are exponentially small in system size. Clearly, the fault-tolerant nature of the quantum memory is intimately related to the existence of topological order in the ground state, or at $T=0$.

At any nonzero temperature, the 2d toric code is not topologically ordered~\cite{Hastings2011}. Heuristically, this is because at any $T>0$, the anyons exist at some finite density. As the system reaches thermodynamic equilibrium, these anyons wander along paths than can be large compared to the system size, connecting the different ground states. The result is that there is only a single equilibrium thermal state, and it does not possess topological order. This suggests that the 2d toric code can not store quantum information indefinitely at $T>0$, but the lack of topological order is an equilibrium property, while any quantum memory properties must be dynamical. 

We will discuss the dynamics of quantum systems evolving at nonzero temperatures following the conventions of Ref.~\cite{RobertsBartlett}. To model the evolution of a system with Hamiltonian $H_\text{sys}$ at some nonzero temperature, we evolve with the full Hamiltonian
\begin{align}
H_\text{full} = H_\sys + H_\bath + \lambda \sum_\alpha S_\alpha \otimes B_\alpha, \label{eqn:bath}
\end{align}
where $H_\bath$ is the bath Hamiltonian. The index $\alpha$ runs over local operators on the system $S_\alpha$ with some corresponding operators on the bath $B_\alpha$. 

When a thermal bath disorders a memory, it does so by applying a logical operator. From Eqn.~\ref{eqn:bath}, this happens when some product of $S_\alpha$ form a logical operator. Thus, the bath can only apply logical operators transversally, or as a series of local operators, each of which is an incomplete logical operator. Incomplete logical operators always anticommute with some terms in the Hamiltonian, so the transveral application of a logical operator must overcome some energy barrier. 

It turns out that the 2d toric code cannot store quantum information indefinitely at $T>0$ without active correction~\cite{Dennis2002}. As with the topological order, the problem is that the point-like anyons exist at finite density at finite temperature, and can wander across the system.

The 4d toric code~\cite{Dennis2002} is analogous to~\eqref{eq:HTC} but in 4 dimensions with qubits on faces, $A_e$ terms on edges, and $B_c$ terms on cubes. 
It evades the issues with finite densities of anyons because the logical operators live on membranes that stretch across the whole system. The topologically-charged excitations, which now live on the boundaries of incomplete logical membranes, are loop-like. A finite temperature bath can create loop excitations of any finite size, but larger loops are suppressed by having larger energy. As the system size $L$ increases, the time that we have to wait for the bath to create $L$-sized loops increases without bound. In fact, the system prefers to shrink any loops that do exist in order to lower the energy. A system that is able to correct errors generated by the bath in this sense is  self-correcting.

In the thermodynamic limit, the bath never creates loops that are as large as the system, so quantum information can be stored indefinitely. In addition, the 4d toric code does remain topologically ordered for nonzero temperatures up to a critical temperature $T_\text{c}$. Above $T_\text{c}$ the 4d toric code is also no longer self-correcting. Thus, self-correction appears to be related to topological order at $T>0$ in the same sense that fault tolerance is related to topological order at $T=0$. The 4d toric code is self-correcting and possesses topological order below $T_\text{c}$. 
Sadly, our world only has 3 spatial dimensions, so we would like to reproduce this behavior in a 3d system.

A feature that distinguishes the 4d toric code from the 2d toric code is that it has an unbounded energy barrier: applying any logical operation through a series of local operation requires traversing a high-energy state, whose energy continues to increase for larger system sizes. The energy of such a state is called the energy barrier of the logical operator. One might be tempted to draw the conclusion that an unbounded energy barrier is sufficient for self-correction. This seems reasonable because operations that cost a divergent energy $\Delta$ should only occur on timescales $\tau \sim \exp (\beta \Delta)$, which is called the Arrhenius law. 
Any string-like logical operator will have a bounded energy barrier because, once the endpoints of the string are well-separated, each endpoint becomes a point-like excitation with constant energy. Thus, the search for models with unbounded energy barriers reduces to a search for models free from string-like logical operators.

In fact, while it is possible to construct 3d systems where all energy barriers are unbounded~\cite{HaahCode, MichnickiPowerLaw},
even these systems do not perform self-correction~\cite{Siva2017}. As in the 2d toric code, the problem can be traced to the existence of topologically-charged point-like excitations~\cite{PremHaahNandkishore}. At nonzero temperature, these excitations exist at finite density. Then, on a very heuristic level, the bath only needs to transport each topological excitation a finite distance to its nearest neighbor. Since the timescale for these partial logical operators is finite, the bath can perform logical operations in a finite time. 

The conclusion to draw here is that unbounded energy barriers are necessary but not sufficient for self-correction. On the other hand, local thermal baths cannot apply membrane-like operators in any finite time, in the thermodynamic limit below some critical temperature~\cite{Dennis2002, RobertsBartlett}, so we expect a memory wherein all logical operators are membrane-like will be self-correcting. As an example, the logical operators in the 4d toric code are all membrane-like.

Instead of looking for a quantum memory that is self-correcting under its own dynamics, we can imagine coupling a toric code to another system in such a way that the latter endows the former with long-range interactions, confining the anyons. For simplicity, assume the coupled system consists of bosons. The 2d version of this proposal is the toric-boson model~\cite{Hamma2009}. The toric-boson model requires couplings between anyons and bosons to have a divergent energy scale. Furthermore, the dynamics of the bosons must be fine-tuned so they do not develop a gap. The 3d version~\cite{Pedrocchi2013} drops the requirement of divergent energy scales, but still needs fine-tuned dynamics~\cite{LandonCardinal2015}. A further difficulty of the generalized toric-boson models is that the boson parts have infinite local Hilbert space dimensions, which is not as useful for quantum computing applications~\cite{Brown2016}. In Sec.~\ref{sub:RB} we will see how the Roberts-Bartlett model reproduces similar physics with finite local Hilbert space dimension.

\subsection{Higher-form symmetries} \label{sub:higher}

Before getting to the Roberts-Bartlett model, let us define higher form symmetries. These generalized global symmetries compactly encode the dynamical constraints required for that model.

In the continuum, an ordinary global symmetry is a group of operators that act on the entire $d$-dimensional space of some theory. As a generalization of global symmetries, $p$-form symmetries act on closed, $(d-p)$-dimensional submanifolds of space~\cite{Gaiotto2015, LakeHigher}. In this classification, ordinary global symmetries can be called 0-form symmetries. Unlike gauge symmetries, which are just redundancies in some description of a theory, higher-form global symmetries are physical symmetries that transform between distinct states. They can give rise to symmetry-protected topological phases~\cite{Gaiotto2015} and symmetry-broken phases~\cite{Gaiotto2015, LakeHigher, WenHigher}, like ordinary global symmetries. 

On a lattice, the definition of the higher-form symmetry needs to be clarified. The proper way to do this is in the language of cellular homology~\cite{Qi2021}.  We will instead proceed by example.

It is easy to find higher-form symmetries in topological phases. In fact, spontaneous breaking of higher-form symmetries leads to topological order~\cite{Gaiotto2015, WenHigher}. As an example, the 2d toric code with no perturbations has an $X$-type and a $Z$-type 1-form symmetry, partially generated by the vertex and face terms, respectively.

An arbitrary product of face terms $B_f = \prod_{e \in f} Z_e$ for some set $F$ of faces gives a symmetry operator $W_\cal{C} = \prod_{e \in \cal{C}} Z_e$.\footnote{The notation is meant to reflect that this operator becomes a Wilson operator if we interpret the toric code as a model for a $\mathbb{Z}_2$ gauge theory.} The path $\cal{C} = \partial F$ is a (possibly disconnected) closed path on the lattice. It is closed in the sense that it does not have any endpoints. Since $\cal{C}$ is the boundary of a collection of faces, these symmetry operators are topologically trivial, meaning they do not wrap around the system. We will call these operators the local part of the symmetry, even though the operators may be large.

There are also topologically nontrivial symmetry operators that do wrap around the system. These are the logical operators in the toric code, which are also closed. We will say they are are the topological part of the symmetry. Both types of operators act on $(1=d-1)$-dimensional paths, so they jointly generate the $Z$-type 1-form symmetry. A similar story exists for the $X$-type 1-form symmetry, with symmetry operators $T_{\cal C'} = \prod_{e\in \cal C'}X_e$ where $\cal C'$ is a path on the dual lattice.

In the 3d toric code with qubits on edges~\cite{CastelnovoFiniteTemp}, the terms $B_f = \prod_{e \in f} Z_e$ still act on the four edges around a face, while the terms $A_v = \prod_{e\in\partial^\dag v} X_e$ now act on the six edges around a vertex.
The face terms still generate 1-dimensional symmetry lines. This means that they are part of a 2-form symmetry now. The vertex terms generate membrane-type operators, which are 2-dimensional objects and therefore part of the 1-form symmetry. These membranes are closed in the sense that they do not have any boundaries.

The higher-form symmetries just described are different than the continuum higher-form symmetries usually considered in the high energy literature~\cite{Seiberg2020, Qi2021}. To understand the difference, recall that the toric code is a model for $\Z_2$ gauge theory. If we were really studying gauge theory, we would identify any states related by a gauge transformation as the same physical state. In the toric code, this means requiring that $A_v=1$ hold as an operator equation for all $v$. Thus, the entire local part of the $X$-type 1-form symmetry acts trivially on the physical Hilbert space. Only the topological part of the $X$-type 1-form symmetry acts nontrivially. Furthermore, any two operators that are topologically equivalent (in the same homology class) are equivalent as operators on the physical Hilbert space.

The two ways of defining higher-form symmetries are called faithful and topological, respectively~\cite{Qi2021}. Faithful higher-form symmetries are more natural in lattice models when we do not want to restrict the Hilbert space, and in non-relativistic models. Topological higher-form symmetries are more natural in gauge theories and relativistic theories~\cite{Seiberg2020}.  In this paper we will discuss faithful higher-form symmetries in order to preserve the tensor-product structure of the global Hilbert space.

When we say that we will enforce a symmetry, we mean that we require that the operators $S_\alpha$ that appear in \eqref{eqn:bath} must commute with the generators of the symmetry. Any local operator that fails to commute with a topological generator also fails to commute with a local generator, so it is enough to require that all the $S_\alpha$ commute with the local part of the symmetry. 

\subsection{Self-correction with a 1-form symmetry} \label{sub:RB}

Now that we have defined higher-form symmetries, we can ask the following question: ``Is it possible to construct a self-correcting quantum memory if we allow ourselves to enforce a 1-form symmetry?" At first, this might seem like an interesting question. Ordinary (0-form) SPT phases are not stable at finite temperature because thermal effects can violate the symmetry locally. On the other hand,
1-form symmetry-protected phases are stable at nonzero temperature, essentially because the symmetry imposes stronger constraints~\cite{Roberts2017}. 

We can quickly see that the answer is trivially ``yes". As an example, take the 2d toric code and require that the dynamics respect all vertex and face terms. In that case, the only allowed operators are products of stabilizers or complete logical operators. If we restrict our bath to only be able to apply operators of bounded size, the bath cannot apply any logical operators. Previously, we could have said that the 1-form symmetries were enforced energetically, in the sense that anyons (which break the symmetry) were suppressed by the gap. Now, we can say the symmetry is enforced explicitly rather than energetically.

Similarly, we can consider the 3d toric code with the vertex terms (which generate a 1-form symmetry) enforced, but not the face terms (which would generate a 2-form symmetry). Once again, the string logical operators can not be applied transversally. The interesting difference is that while the membrane operators can be applied transversally without breaking the symmetry, the memory time still grows without bound. This is because of the previous argument that thermal baths cannot apply membrane operators in the thermodynamic limit~\cite{Dennis2002, CastelnovoFiniteTemp}.

A more interesting question to ask is: ``Is it possible to construct a self-correcting quantum memory that permits the transversal application of logical Pauli operators, if we allow ourselves to enforce a 1-form symmetry?" Neither the 2d or 3d toric codes with 1-form symmetry enforced answer this question. Instead, the Roberts-Bartlett model shows that the answer is ``yes"~\cite{RobertsBartlett}, constructing a model that consists of the 3d cluster state Hamiltonian of Raussendorf, Bravyi, and Harrington (RBH)~\cite{Raussendorf2005}, with 2d toric code boundary conditions. When the 1-form symmetry is enforced in the bulk, the boundary logical degrees of freedom do not evolve in time, even at nonzero temperature.

We should note that the Roberts-Bartlett model does not allow for the transversal application of arbitrary (non-Pauli) logical operators. In fact, the same is true of the 4d toric code (which is self-correcting at $T>0$ without any enforced symmetry). There is a model that supports arbitrary transversal logical operations in 7 spatial dimensions, but no such model in fewer dimensions is known~\cite{Bombin2013}.

The explicit construction of the Roberts-Bartlett model is rather involved, so we leave the details to the original literature~\cite{RobertsBartlett}. Here, we will only review the models at the effective level.
The RBH Hamiltonian is not topologically ordered, but is SPT-ordered under a 1-form symmetry~\cite{Roberts2017}. Furthermore, this SPT order is stable at nonzero temperatures. Ordinary (0-form) SPT order does not survive to nonzero temperatures~\cite{Roberts2017}.

The logical information in the Roberts-Barlett model resides on the boundary toric code qubits. As always, anyons live at the ends of partial logical operators.  In the Roberts-Bartlett model, the anyons are connected to extended excitations (flux strings) that extend into the bulk and have linear energy cost. The 1-form symmetry then ensures that these bulk flux strings cannot end, except on another anyon on the boundary.~\cite{RobertsBartlett}.

\begin{figure}
    \centering
    \includegraphics[width=.7\linewidth]{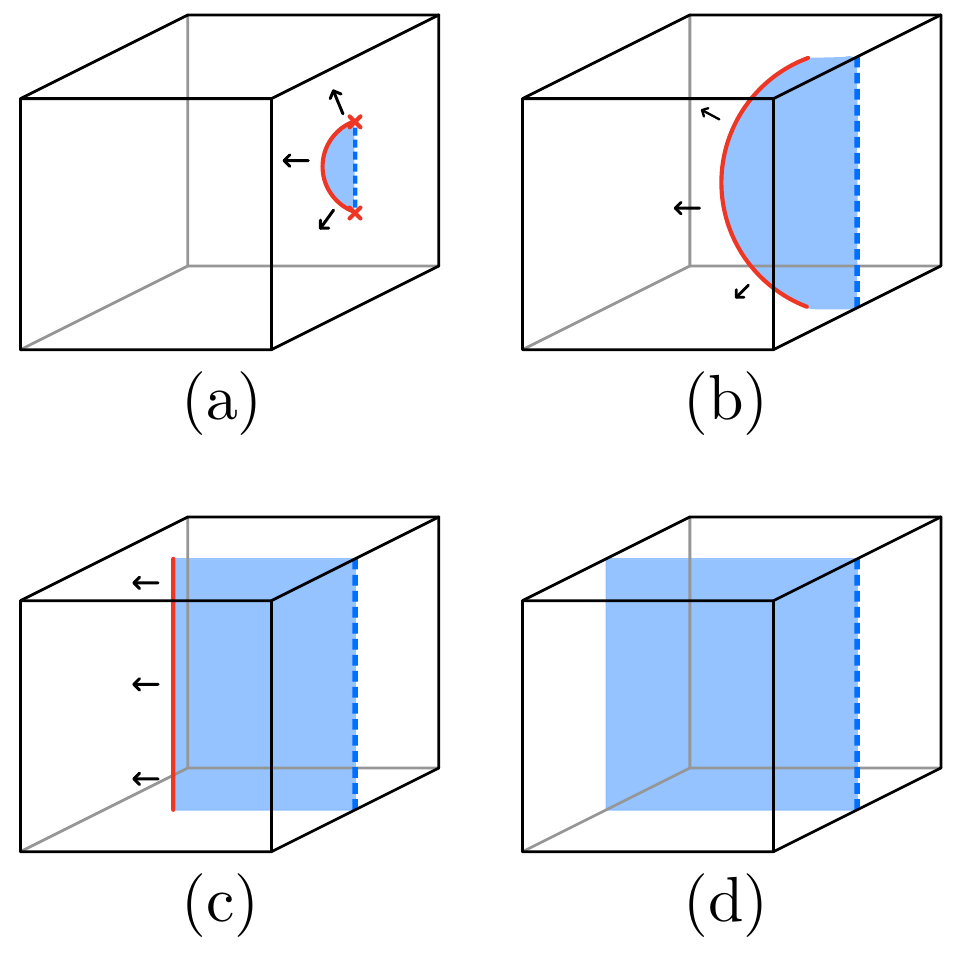}
    \caption{Decomposition of a logical operator. The top and bottom of the cube are identified and the front and back of the cube are identified. In the RBH model and in the symmetry-protected boundary memory introduced in this paper, anyons on the boundary are attached to flux in the bulk. Red crossed represent anyons, red lines are bulk flux, and blue lines are logical operators on the boundary. The blue shaded region is a membrane operator, which is necessary for the symmetry-protected boundary memory but not for the RBH model. Compare to Fig.~12 of~\cite{RobertsBartlett} and Fig.~7 of~\cite{Stahl2021}.}
    \label{fig:flux}
\end{figure}

Once the anyons are connected to the bulk flux, they are confined, and cannot traverse the system through thermal effects. In 2d, confinement ruins topological order because if the anyons leave energetic flux behind, then the full operator cannot be a logical operator (because it does not commute with the Hamiltonian). Instead, the Roberts-Bartlett model uses the third space dimension to remove the energetic flux. Although specific details of this procedure depend on the explicit construction, Fig.~\ref{fig:flux} shows the removal on the effective level.

Reference~\cite{RobertsBartlett} also shows that once the anyons are confined by flux strings with linear energy cost, the memory time of the model will grow without bound in the thermodyanmic limit. This is different from fracton models, where a diverging (but sub-linear) energy barrier does not lead to a diverging memory time~\cite{Siva2017}.

The Roberts-Bartlett construction can be extended to any Walker-Wang model~\cite{WalkerWang, vonKeyserlingkSurfaceAnyons}. In fact, the RBH Hamiltonian (the bulk of the Roberts-Bartlett model) is equivalent to the Walker-Wang model with the toric code braided fusion category as input, after moving some qubits to faces~\cite{Roberts2020}. Furthermore, the Roberts-Bartlett construction can be extended to a model with a trivial bulk,~\cite{Stahl2021} at the cost of enforcing a 1-form symmetry with an action at the boundary that is not on-site~\cite{WenHigher}. 

We can gain a new perspective on the Roberts-Bartlett model by examining the precise role of the 1-form symmetry in protecting the memory. As emphasized in Ref.~\cite{RobertsBartlett}, the 1-form symmetry in the bulk prevents the flux-like excitations from ending. On the boundary, the 1-form symmetry requires that any boundary anyons live on the endpoints of bulk fluxes. Both roles are essential in this family of models. If boundary anyons did not need to be attached to bulk excitations, they would be deconfined. If the bulk fluxes were allowed to end, then boundary anyons could be attached to finite-length bulk fluxes, again leading to deconfinement.

Reference~\cite{RobertsBartlett} already pointed out that topologically-ordered models like the 3d toric code can have an emergent 1-form symmetry, so that there are loop excitations that cannot end in the bulk. 
The contribution of the current paper is to demonstrate that we can use this emergent symmetry to replace the enforced symmetry in the bulk. However, we have not found a way for the emergent symmetry to attach the anyons to the bulk flux. Instead, we will  need to enforce a 1-form symmetry on the boundary. This means that we can achieve self-correction while enforcing an $\cal{O}(L^2)$ number of stabilizers, rather than an $\cal{O}(L^3)$ number.

\section{Restricting to a boundary symmetry} \label{sec:boundary}

In this section we will introduce our new model that uses an emergent 1-form symmetry in the bulk rather than directly enforcing a bulk symmetry. As the model consists of a memory on a boundary protected by a symmetry, we will call it the symmetry-protected boundary memory. We will first define the Hamiltonian and then describe the logical operators. We will then explain why a subspace of the logical codespace forms a memory that is stable at finite temperature.

\subsection{Hamiltonian and logical operators} \label{sub:Hamiltonian}

\begin{figure}[htbp]
\centering
\includegraphics[width=.7\linewidth]{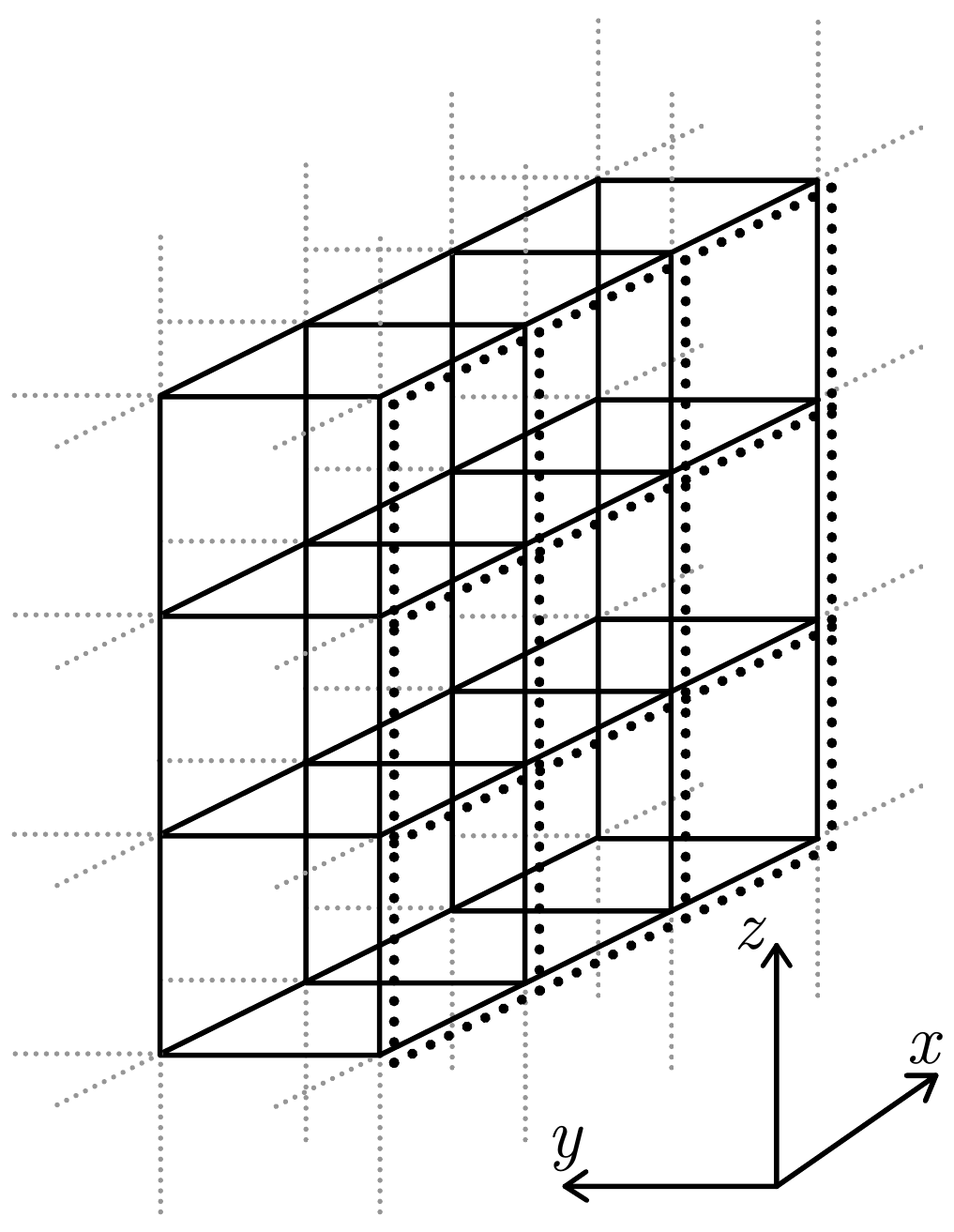}
\caption{We can view the Hilbert space of the model as living on a cubic lattice. In the bulk, qubits live on faces and edges. The boundary faces have no qubits. The boundary edges have two qubits each, one of which is a bulk $(e)$ degree of freedom (solid line) and one of which is a boundary $(b)$ degree of freedom (dotted line). The lattice is periodic in the $x$ and $z$ directions and has boundaries at $y=0$ and $y=L$. This figure shows the ``right" boundary at $y=0$, where we will put our 2d toric code. The boundary at $y=L$ is gapped with no additional topological order.}
\label{fig:hilbert}
\end{figure}

In the absence of the symmetry, the symmetry-protected boundary memory is a tensor product of two 3d toric codes with a 2d toric code on their boundary. It lives on a cubic lattice with boundary. 
The Hilbert space is as in Fig.~\ref{fig:hilbert}, with qubits on edges and faces and a second qubit on each boundary edge. Thus, we have enough qubits to define two copies of the 3d toric code and one copy of the 2d toric code. The full Hamiltonian will be
\begin{align}
H &= H^{(e)} + H^{(f)} + H^{(b)}, \label{eq:fullH}
\end{align}
with each term defined below. We will use a lattice that is periodic in the $z$ and $x$ directions, so that the global structure is a thickened torus, $T^2\times I$. There are two boundaries, each a 2d torus, with the 2d toric code on the $y=0$ boundary as shown in Fig.~\ref{fig:hilbert}. The other boundary (at $y=L$) will have similar boundary conditions on the $(e)$ and $(f)$ qubits, but without the extra boundary degrees of freedom.

\begin{figure}
\centering
\includegraphics[width=\linewidth]{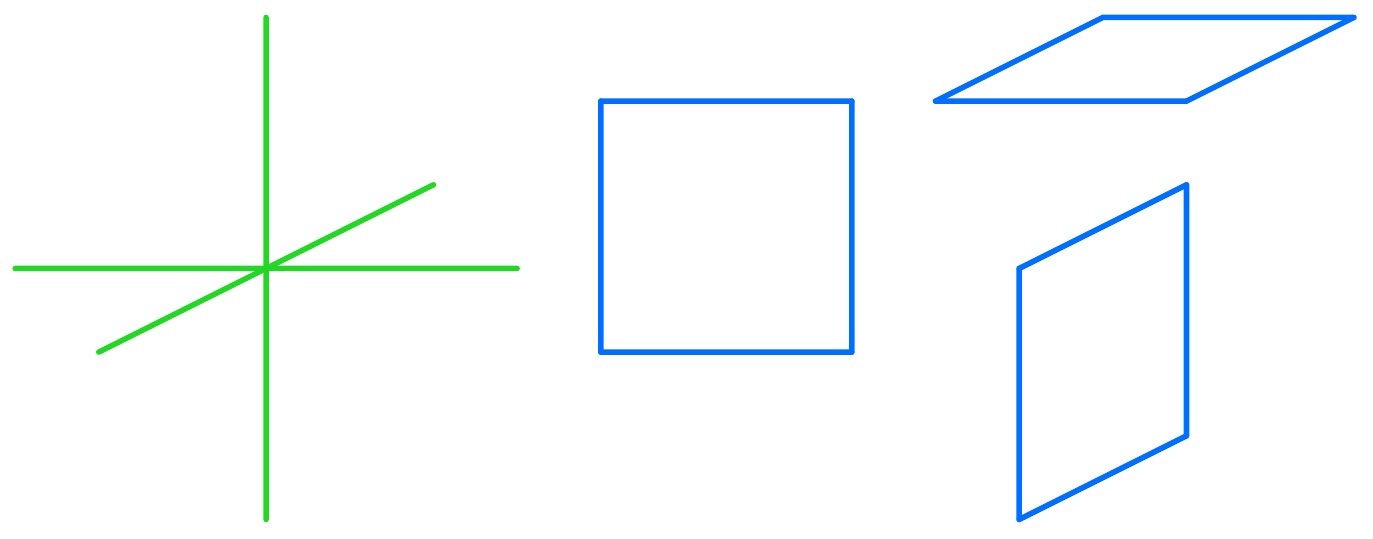}
\caption{Stabilizers for the 3d toric code on edges. Green edges are $X^{(e)}$ operators and blue edges are $Z^{(e)}$ operators.}
\label{fig:edges}
\end{figure}

The first term in $H$,
\begin{align}
H^{(e)} &= -\sum_vA_v^{(e)} - \sum_f B_f^{(e)},
\end{align}
is a bulk 3d toric code Hamiltonian acting on edge degrees of freedom. The individual stabilizers,  
\begin{align}
A_v^{(e)} = \prod_{e\in\pardag v}X_e^{(e)},\qquad B_f^{(e)} = \prod_{e\in\partial f} Z_e^{(e)}
\end{align}
are shown in Fig.~\ref{fig:edges}.  The logical operators in this sector are direct lines of $Z^{(e)}$ operators and dual membranes\footnote{A dual membrane is a membrane on the dual lattice. The dual lattice is the result of exchanging vertices with cubes and interchanging edges with faces.} of $X^{(e)}$ operators. At endpoints of $Z^{(e)}$ lines we have $e^{(e)}$ anyons and on the boundaries of $X^{(e)}$ membranes we have $m^{(e)}$ flux. The boundary conditions at $\pm y$ are ``smooth", so that the $m^{(e)}$ flux is condensed (meaning the flux can be removed at the boundary) and $e^{(e)}$ anyons are confined. Equivalently, $X^{(e)}$ membranes can terminate on the boundaries but $Z^{(e)}$ lines cannot.

\begin{figure}
\centering
\includegraphics[width=\linewidth]{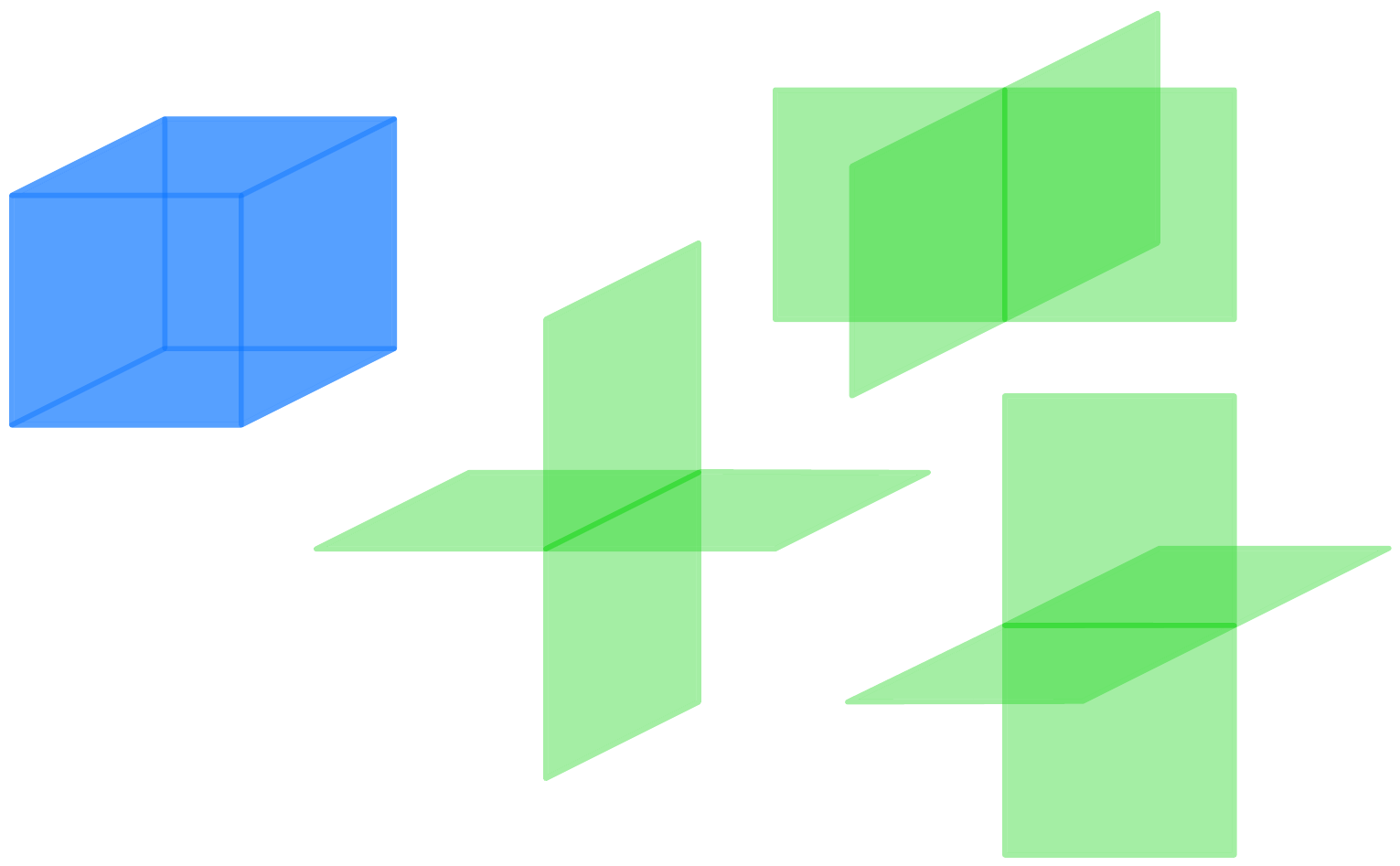}
\caption{Stabilizers for the 3d toric code on faces. Green faces are $X^{(f)}$ operators and blue faces are $Z^{(f)}$ operators.}
\label{fig:faces}
\end{figure}

The other bulk Hamiltonian is 
\begin{align}
H^{(f)} &= -\sum_cA_c^{(f)} - \sum_e B_e^{(f)},
\end{align}
which acts on face qubits. The terms are 
\begin{align}
A_c^{(f)} = \prod_{f\in\partial c}Z_f^{(f)},\qquad B_e^{(f)} = \prod_{f\in\pardag e} Z_f^{(f)},
\end{align}
as shown in Fig.~\ref{fig:faces}. They are equivalent to the edge terms after swapping edges with faces, swapping cubes with vertices, and swapping $X$ with $Z$. The logical operators in this sector are dual lines of $X^{(f)}$ operators and direct membranes of $Z^{(f)}$ operators. Here, the excitations are point-like $m^{(f)}$ anyons on the ends of $X^{(f)}$ dual lines and extended $e^{(f)}$ flux at the boundaries of $Z^{(f)}$ membranes. Note that we are using a convention for the $(f)$ degrees of freedom where the electric excitations are extended and the magnetic excitations are point-like. The boundary conditions are such that the $m^{(f)}$ flux is condensed, which is now called the ``rough" boundary conditions.

\begin{figure}
\centering
\includegraphics[width=.6\linewidth]{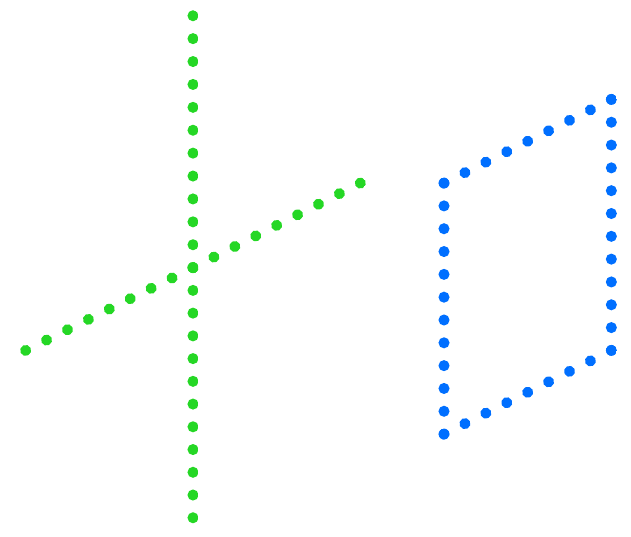}
\caption{Stabilizers for the 2d toric code on the boundary. Green dotted lines are $X^{(b)}$ operators and blue dotted lines are $Z^{(b)}$ operators.}
\label{fig:boundary}
\end{figure}

On the extra set of boundary qubits we will define a 2d toric code,
\begin{align}
\begin{aligned}
H^{(b)} &= -\sum_vA_v^{(b)} - \sum_f B_f^{(b)},\\
A_v^{(b)} &= \prod_{e\in\pardag v}X_e^{(b)},\qquad B_f^{(b)} = \prod_{e\in\partial f} Z_e^{(b)},
\end{aligned}
\end{align}
where the sums and products are only taken over boundary faces, edges, and vertices. These terms are shown in Fig.~\ref{fig:boundary}. Here, the logical operators are direct lines of $Z^{(b)}$ and dual lines of $X^{(b)}$. The excitations are $e^{(b)}$ anyons and $m^{(b)}$ anyons.

For simplicity let us only describe a subset of the logical qubits, and therefore a subset of the logical operators. Let $\Bar{Z}^{(e)}$ correspond to a vertical line of $Z^{(e)}$ operators in the bulk and $\bar{X}^{(e)}$ to a horizontal dual membrane of $X^{(e)}$ operators (which must intersect both boundaries). For the face code let $\bar{Z}^{(f)}$ be a vertical membrane of $Z^{(f)}$ operators and $\bar{X}^{(f)}$ a horizontal dual line through the bulk of $X^{(f)}$ operators. On the boundary we have $\bar{Z}^{(b)}$ a vertical line of $Z^{(b)}$ operators and $\bar{X}^{(b)}$ a horizontal dual line of $X^{(b)}$ operators.  There are another 3 logical qubits whose logical operators are related by a global rotation, but we can safely ignore these as their description is the same. 

All three of these qubits are fault tolerant. This corresponds to the existence of topological order at $T=0$~\cite{Kitaev2003}. At nonzero temperature, a local thermal bath can apply the operators $\Bar{Z}^{(e)}$, $\bar{X}^{(f)}$, $\bar{Z}^{(b)}$, and $\bar{X}^{(b)}$, essentially because they have a constant energy barrier. Thus the edge and face logical qubits can serve as classical, but not quantum, memories~\cite{CastelnovoFiniteTemp}, while the boundary logical qubit can store no information.

\subsection{Enforcing the symmetry} \label{sub:symm}

\begin{figure}
\centering
\includegraphics[width=.6\linewidth]{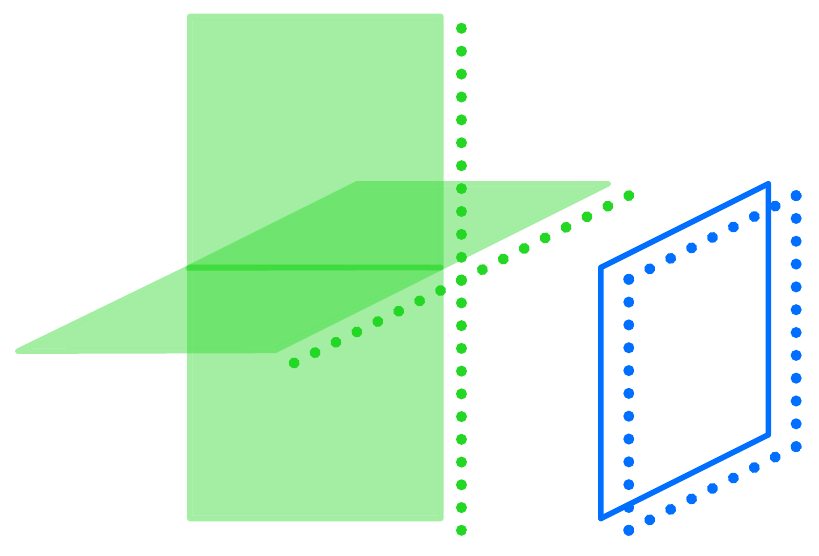}
\caption{The terms we enforce are $\cal{A}_v = A_v^{(b)} A_{e^{(0)}}^{(f)}$ (left) and $\cal{B}_f = B_f^{(e)} B_f^{(b)}$ (right). Since the symmetry operators are products of stabilizers, the ground space is not affected. Instead, some excitations are forbidden, so that boundary logical operators can only be applied in tandem with bulk membrane operators.}
\label{fig:enforce}
\end{figure}

The last ingredient is the 1-form symmetry that we will choose to enforce. The symmetry acts on the boundary, in the sense that it acts only on $(b)$ qubits, and those $(e)$ and $(f)$ qubits that are adjacent to the boundary. For $v$ a boundary vertex and $f$ a boundary face, the generators of the local part of the symmetry,
\begin{align}
\cal{A}_v = A_v^{(b)} A_{e^{(0)}}^{(f)}, \qquad \cal{B}_f = B_f^{(b)}B_f^{(e)}, \label{eqn:sym}
\end{align}
are products of stabilizers in Hamiltonians. The edge $e^{(0)}$ is the unique non-boundary edge such that $v\in\partial e^{(0)}$.  The terms are illustrated in Fig.~\ref{fig:enforce}. The generators of the topological part of the symmetry are products of logical operators, such as $\bar{Z}^{(e)}\bar{Z}^{(b)}$. The local and topological generators locally look the same.

\begin{figure}
\centering
\includegraphics[width=.6\linewidth]{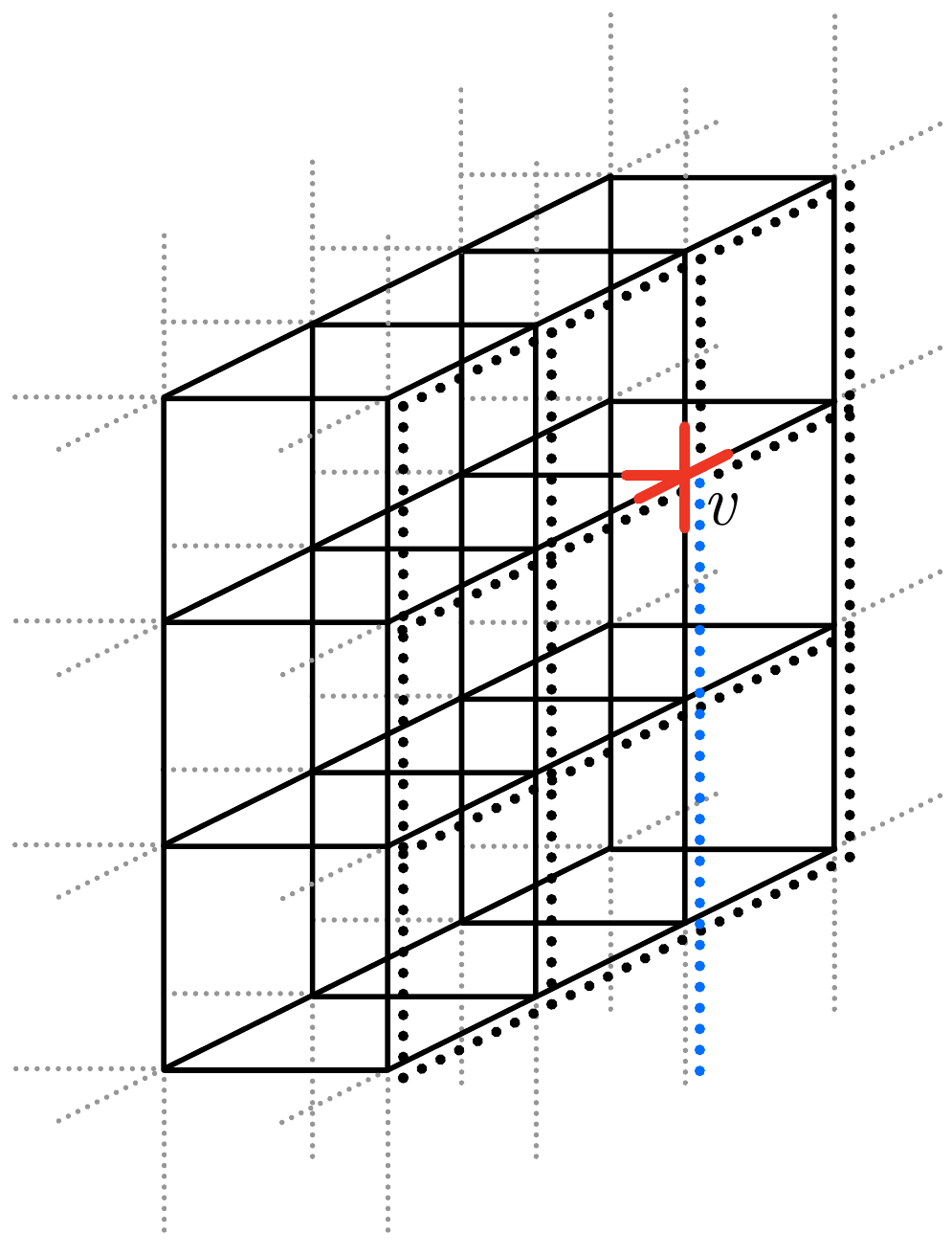}
\caption{Blue dotted lines represent $Z^{(e)}$ operators, a partial application of the logical operator $\bar{Z}^{(e)}$. At the highlighted vertex, the partial logical operator anticommutes with $A_v^{(b)}$ and $\cal{A}_v$. Anticommutation with $A_v^{(b)}$ only leads to an energy penalty, but anticommutation with $\cal{A}_v$ means that this operator is forbidden by the 1-form symmetry. A similar argument applies to any partial $\bar{Z}^{(e)}$ operator.}
\label{fig:forbidden}
\end{figure}

When the symmetry is enforced, $e^{(b)}$ anyons are required to coincide with endpoints of $e^{(f)}$ flux and $m^{(b)}$ anyons are required to coincide with endpoints of $m^{(e)}$ flux, all of which occur only on the lattice boundary. This means that the operators $\bar{Z}^{(b)}$, $\bar{Z}^{(f)}$, $\bar{X}^{(b)}$, and $\bar{X}^{(e)}$ all can no longer be applied transversally.  For the boundary operators this is because the boundary anyons are prohibited. This is demonstrated for $\bar{Z}^{(e)}$ in Fig.~\ref{fig:forbidden}. For the two membrane operators, open membranes (incomplete logical operators) are permitted in the bulk but  are not allowed to intersect the boundary. The bulk line operators $\bar{Z}^{(e)}$ and $\bar{X}^{(f)}$ can still be applied transversally because they need not intersect the boundary. 

\begin{figure}
\centering
\includegraphics[width=.6\linewidth]{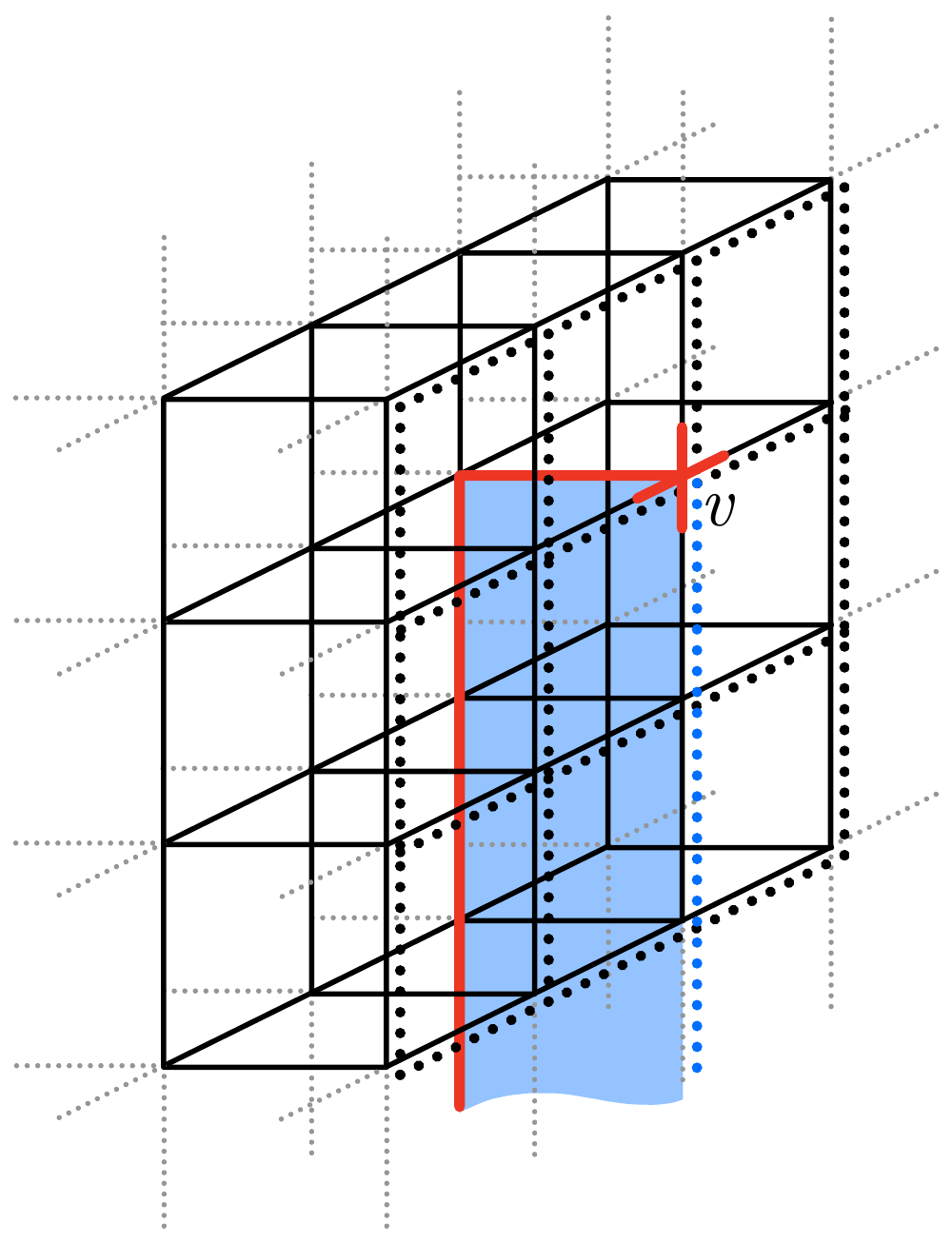}
\caption{Blue dotted lines represent $Z^{(e)}$ operators and blue faces are $Z^{(f)}$ operators. This operator anticommutes with $A_v^{(b)}$ at the highlighted vertex and $B_e^{(f)}$ at every highlighted edge, which leads to a large energy penalty. The addition of the $Z^{(f)}$ operators means it now commutes with $\cal{A}_v$, so it is a permitted operator. We should view this as a partial application of the composite logical operator $\bar{Z}^{(f)} \bar{Z}^{(b)}$.}
\label{fig:composite}
\end{figure}

The only way to transversally act on the boundary logical qubit is through the composite logical operators $\bar{X}^{(e)} \bar{X}^{(b)}$ and $\bar{Z}^{(f)} \bar{Z}^{(b)}$. Figure~\ref{fig:composite} demonstrates the partial application of $\bar{Z}^{(f)} \bar{Z}^{(b)}$. Since both composite operators include a membrane part, the composite operators are linearly confined. The upshot is that all logical operators that can transversally act on the boundary logical qubit are linearly confined. We should notes that while we need the bulk topological order to supply the bulk flux, we do not store any information in the bulk.

Since both boundary logical operators are linearly confined, a local bath cannot apply them in finite time in the thermodyamic limit~\cite{RobertsBartlett}. This means that the current model achieves the same memory properties as the Roberts-Bartlett model, while only requiring that a symmetry be enforced at the boundary. No symmetry terms need to be enforced in the bulk.

\section{Interpretation} \label{sec:interp}

We can gain some insight into the behavior of this model by analyzing it on the effective level, instead of focusing on the specific lattice realization. There, the simplest language to use is that of defects in 3d topological orders. So far, we have been orienting the model so that the 2d toric code sits on the boundary of the system. Since the edge and face bulk degrees of freedom are noninteracting, we can alternatively unfold~\cite{Kitaev2012} the two bulks and view the 2d toric code as a boundary or \emph{defect} between two spatially separated 3d toric codes. Reference~\cite{Aasen2020} explains defects in depth, furthermore using networks of defects (and defects of defects, etc.) to construct fracton phases. Here, we will only need to discuss 2d defects in 3d topological orders. 

We will describe defects as the result of a process in which we confine and condense composite objects at a 2d surface in a 3d topological order. Condensation of composite objects can be useful in constructing many models, including Michnicki's welded code~\cite{MichnickiPowerLaw, Siva2017} and fractons~\cite{Ma2017, Vijay2017, Qi2021}. When a condensed composite $a_1\dots a_n$ has nontrivial mutual statistics with another anyon $a'$, the anyon $a'$ becomes confined. Similarly, when the same composite is condensed on a boundary, $a'$ becomes confined on that boundary, meaning it cannot be annihilated at and cannot pass through that boundary.

Let us say that a flux string is deconfined on a certain boundary if it can end on that boundary, and confined if it can not. Under this definition, in the 3d toric code (defined on edges), flux strings are confined at the rough boundary and deconfined at the smooth boundary. Furthermore, the $e$ anyons are confined at the smooth boundary and condensed on the rough boundary. If we take a smooth boundary and condense $e$ anyons, the flux strings become confined and we end up with a rough boundary. This generalizes the previous notion of condensation leading to confinement. If we start with a rough boundary and condense $m$ flux near the boundary, the $e$ anyons become confined on that boundary.

Now recall from Sec.~\ref{sub:RB} that a 2d toric code with both 1-form symmetries enforced is trivially self-correcting because it has no dynamics. We can view enforcing the 1-form symmetry as confining the $m$ and $e$ anyons ``by hand," or without any condensation procedure. 

\begin{figure}
    \centering
    \includegraphics{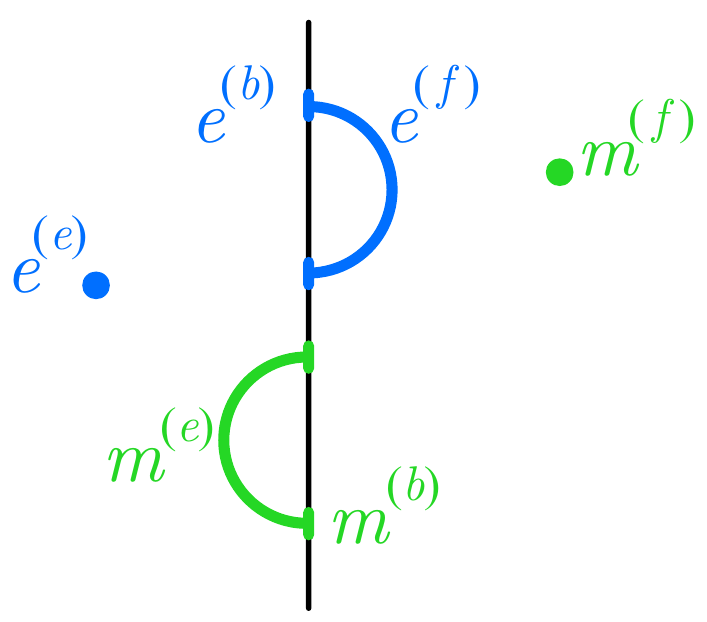}
    \caption{The setup for the symmetry-protected boundary memory has $e^{(e)}$ and $m^{(f)}$ anyons and $m^{(e)}$ and $e^{(f)}$ flux in the two bulks, and $e^{(b)}$ and $m^{(b)}$ anyons on the defect. For the microscopic realization in Sec.~\ref{sec:boundary}, the green excitations are created from $Z$-type errors while the blue excitations are created from $X$-type errors. The confinement procedure ensures that $e^{(b)}$ anyons coincide with endpoints of $e^{(f)}$ flux and $m^{(b)}$ anyons coincide with endpoints of $m^{(e)}$ flux.}
    \label{fig:defect}
\end{figure}

We can also construct the symmetry-protected boundary memory using by-hand confinement. As in Fig.~\ref{fig:defect}, we have a 3d toric code labeled by $(e)$, a 2d toric code in the center labeled by $(b)$, and another 3d toric code on the left labeled by $(f)$. The labels are chosen to match the labels in Sec.~\ref{sec:boundary}, but we no longer need to refer to edges and faces. On the boundary, we confine the $e^{(b)}$ and $m^{(b)}$ anyons and the $m^{(e)}$ and $e^{(f)}$ fluxes, in such a way that the composite objects $m^{(e)}m^{(b)}$ and $e^{(b)}e^{(f)}$ are deconfined. When we say that a composite of a boundary anyon and a bulk flux are deconfined, we mean that the flux may end on the boundary, but only if its endpoint coincides with the corresponding anyon.  Similarly, the boundary anyons my move freely only when attached to bulk flux.

The bulk fluxes give the boundary anyons dynamics, so the anyons are (linearly) energetically confined, instead of exactly confined like in the trivial 2d toric code example.
As in Sec.~\ref{sec:boundary}, the boundary anyon confinement means that the bath cannot apply logical operators, even though the operators can be transversally applied. Just like in Sec.~\ref{sec:boundary}, though, these boundary conditions need the presence of the 1-form symmetry in order to be stable to perturbations. We can view this as saying that by-hand confinement without condensation is fine-tuned.

Note that, for example, the confinement of $m^{(e)}$ and $m^{(b)}$ but not $m^{(e)}m^{(b)}$ is the confinement pattern that would result from condensing the composite $e^{(e)}e^{(b)}$. In fact, we can follow that condensation procedure on the microsopic lattice by considering the Hamiltonian~\eqref{eq:fullH} as a perturbation to the condensing Hamiltonian 
\begin{align}
H_\text{cond} = -J_x \sum_{e}X_e^{(e)}X_e^{(b)},
\end{align}
where the sum is taken over boundary edges, in the large $J_x$ limit. Then we get the symmetry term $\cal{B}_f$ in~\eqref{eqn:sym} at some order in perturbation theory, which confines $m^{(e)}$ and $m^{(b)}$ but not $m^{(e)}m^{(b)}$. Unfortunately, fully building the symmetry-protected boundary memory from condensation would also require condensing $m^{(b)}m^{(f)}$, which is not possible because condensed composites cannot have mutual statistics. 

Before concluding, let us briefly mention an interesting related model, which can be constructed from a 2d $\Z_4$ topologically-ordered defect in a 3d $\Z_2$ topologically-ordered bulk. The boundary topological order contains anyons $e^{(b)}$, $e^{2(b)}$, $e^{3(b)}$, $m^{(b)}$, $m^{2(b)}$, and $m^{3(b)}$, along with their products. All are abelian, and the $e^{(b)}$ and $m^{(b)}$ anyons have mutual statistics $i$. The bulks are as before. 

On the defect, condense the composite objects $e^{(e)}e^{2(b)}$ and $m^{2(b)}m^{(f)}$. This condensation is consistent because $e^{2(b)}$ and $m^{2(b)}$ have trivial mutual statistics. The procedure also removes all topological order from the defect. The anyons $e^{2(b)}$ and $m^{2(b)}$ are still deconfined, but are respectively equivalent to the $e^{(e)}$ and $m^{(f)}$ bulk anyons. Furthermore, the $m^{(e)}$ flux string is allowed to terminate on the boundary, but only in the presence of a $m^{(b)}$ anyon which is otherwise confined. A similar story exists for the $e^{(f)}$ flux and $e^{(b)}$ anyon. 

\begin{figure}
    \centering
    \includegraphics[width=.6\linewidth]{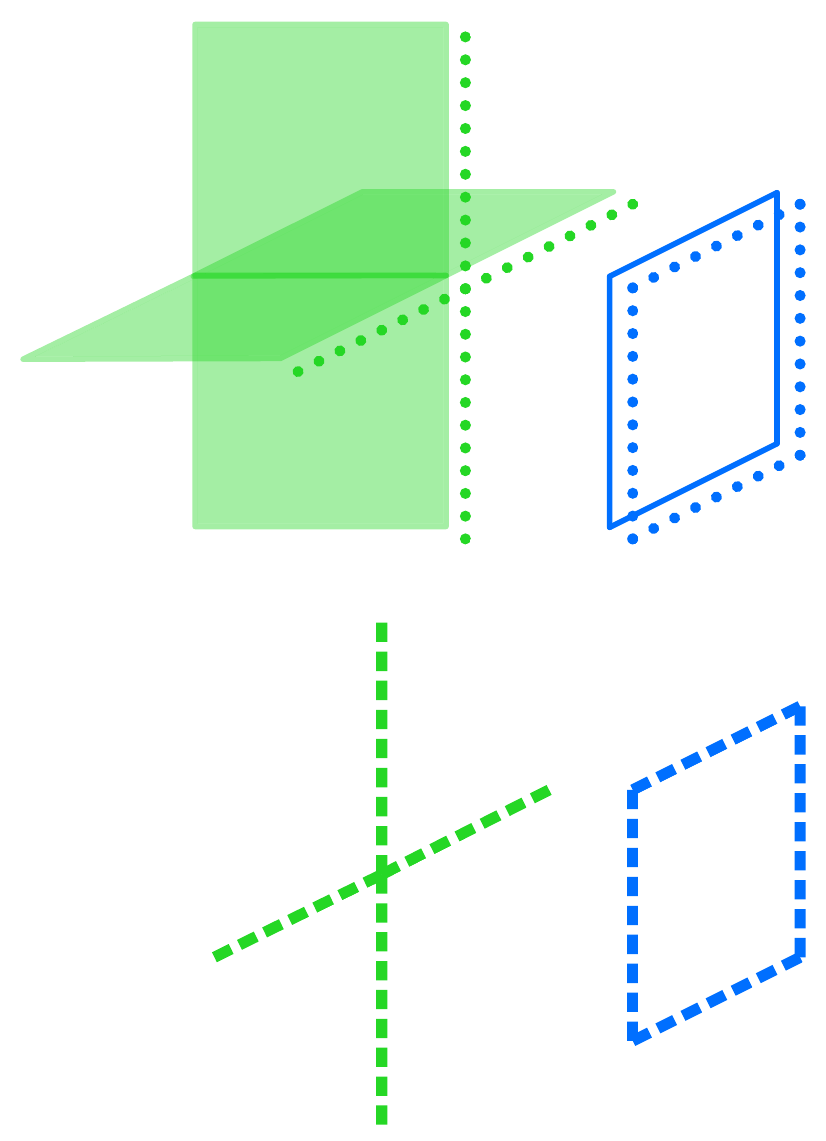}
    \caption{The boundary Hamiltonian terms in \eqref{eqn:Z4bdy}. The top row shows $\mathcal{A}_v$ and $\mathcal{B}_f$, where green faces are $X^{(f)}$, green dotted lines are $X^{(b)}$, blue solid lines are $Z^{(e)}$, and blue dotted lines are $Z^{(b)}$. The bottom row shows $A^{(b)}_v$ and $B^{(b)}_f$, where green dashed lines are now $\big( X^{(b)}_e \big)^2$ and blue dashed lines are $\big( Z^{(b)}_e \big)^2$. The top two terms square to the bottom two, respectively.}
    \label{fig:Z4bdy}
\end{figure}

The resulting model is stable to any small perturbation and does not require any symmetry-protection. We can construct a microscopic model, as in Sec.~\ref{sec:boundary}, by putting qubits on bulk edges and faces, and 4-level qudits on boundary edges. The 4-level qudits have Pauli-like operators obeying
\begin{align}
X^{(b)}_e Z^{(b)}_e = i Z^{(b)}_e X^{(b)}_e, \quad \big( X^{(b)}_e \big)^4  = \big( Z^{(b)}_e \big)^4 = 1,
\end{align}
while the bulk face and edge qubits have ordinary 2-level Pauli operators.
The Hamiltonian consists of the ordinary bulk toric code Hamiltonian, with the boundary Hamiltonian 
\begin{align}
H_\text{bdy} &= -\sum_v \mathcal{A}_v - \sum_f \mathcal{B}_f - \sum_v A_v^{(b)} - \sum_f B_f^{(b)} + \text{h.c.}, \nonumber\\
\mathcal{A}_v &= \prod_{e\in \partial^\dag v} X^{(b)}_e X^{(f)}_{f^{(0)}}, \qquad \mathcal{B}_f = \prod_{e \in \partial f} Z^{(b)}_e Z^{(e)}_e,\nonumber\\
A_v^{(b)} &= \prod_{e\in \partial^\dag v} \big( X^{(b)}_e \big)^2, \qquad B_f^{(b)} = \prod_{e\in \partial f} \big( Z^{(b)}_e \big)^2, \label{eqn:Z4bdy}
\end{align}
where $f^{(0)}$ is the unique non-boundary face in $\partial^\dag e$ and, analogous to in Ref.~\cite{Kitaev2003}, each edge and plaquette must be assigned an orientation.
The sums are over boundary vertices and faces, and the second two terms are just the squares of the first two terms. The terms are shown in Fig.~\ref{fig:Z4bdy} using the ``folded" conventions from Sec.~\ref{sec:boundary}.  

This $\Z_4$ defect model is not self-correcting. At nonzero temperature it can only store a number of probabilistic bits, just like decoupled 3d toric codes~\cite{CastelnovoFiniteTemp}. Instead, it is interesting because membrane operators in the two bulks fail to commute, due to the fact that the bulk fluxes must terminate in anyons on the defect and the respective anyons have nontrivial mutual statistics. The presence of membrane operators that fail to commute seems to be a necessary (although clearly not sufficient) ingredient for self-correction~\cite{Dennis2002, RobertsBartlett}. The $\mathbb{Z}_4$ defect model appears to be the first model in three or fewer dimensions with this property.

\section{Conclusions} \label{sec:conc}

The symmetry-protected boundary memory is useful in two ways. First, it improves upon the Roberts-Bartlett model by only requiring that $\cal{O}(L^2)$ symmetry generators be enforced, rather than $\cal{O}(L^3)$. In that sense it is a continuation of the work in Ref.~\cite{Stahl2021}, which showed that self-correction is possible with a number of enforced terms that is asymptotically smaller than $L^3$ but greater than $L^2$.

The second contribution of the present model is to emphasize that, in the Roberts-Bartlett model, the 1-form symmetry serves two distinct purposes. Namely, it ensures that the flux strings do not end in the bulk \emph{and} requires that boundary anyons and flux string endpoints coincide. Here, we show that the first contribution can be supplied by bulk topological order. Even at nonzero temperature, the flux strings of the 3d toric code cannot end. This is connected to the fact that discrete 1-form symmetries can be spontaneously broken (and therefore emergent~\cite{WenHigher}) at non-zero temperature in 3d.

On the other hand, we have not yet found a way to require that boundary anyons and flux string endpoints coincide without explicitly enforcing the 1-form symmetry at the boundary. Finding a way to make this requirement emergent, rather than explicit, would certainly be exciting. 

It might also be interesting to extend the constructions in this paper to more general boundaries and bulks. For example, generalizations of the symmetry-protected boundary memory should be able to realize more general $G$-crossed braided tensor categories~\cite{Barkeshli2019}. Relatives of the $\Z_4$ defect model could have more interesting string-nets on the boundary, and other exotic phases in the bulk.

Finally, the finite-temperature behavior of topological defects may warrant further exploration. It may be the case that they have no memory properties beyond those of the topologically-ordered bulks, but the existence of noncommuting membrane operators in 3d seems intriguing. 

{\bf Acknowledgements:} This work was supported by the U.S. Department of Energy, Office of Science, Basic Energy Sciences, under Award No. DE-SC0021346.

\bibliography{bob}
	
\end{document}